\documentclass[12pt]{iopart}

\usepackage{iopams}

\begin{document}

\title{Boundary effects of electromagnetic vacuum fluctuations on charged particles}

\author{Tai-Hung Wu, Jen-Tsung Hsiang, Da-Shin Lee}
\address{Department of Physics, National Dong Hwa University,
Hualien 974, Taiwan, R.O.C.}

\begin{abstract}
The effects of electromagnetic vacuum fluctuations with the boundary on charged particles is investigated. They may be observed via an electron interference experiment near the conducting plate, where boundary effects of vacuum fluctuations are found significant on coherence reduction of the electrons. The dynamics of the charge under the influence of quantized electromagnetic fields with a conducting plate is also studied. The corresponding stochastic equation of motion is derived in the semiclassical approximation, and the behavior of the charge's velocity fluctuations is discussed.
\end{abstract}

\maketitle

\section{Introduction}
Manipulation of zero-point fluctuations due to the imposition of the boundary
conditions may lead to an observable impact on macroscopic physics. One of the
most celebrated examples is the attractive Casimir force between two
parallel conducting plates~\cite{CA1}. Nevertheless, this
induced-force effect can also be probed through the coupling to a
test particle. Consider an atom in its ground state as a test
particle located near a perfectly conducting plate. The boundary
induced effects from electromagnetic vacuum fluctuations thus leads
to a position-dependent energy shift, which further results in an
attractive Casimir-Polder force on the atom toward the
plate~\cite{CA2}. Thus, in this paper the motion of the test
particle can serve as a probe to understand the nature of
fluctuations from the effects on its dynamics. The dynamics of this
particle and field interaction has been studied quantum-mechanically
in the system-plus-environment approach~\cite{CAL1,CAL2}. We
treat the particle as the system of interest, and the degrees of
freedom of fields as the environment. The influence of fields on the
particle can be investigated with the method of Feynman-Vernon
influence functional by integrating out field variables within the
context of the closed-time-path formalism ~\cite{FE,SC}. The aim of
this paper is to present a field-theoretic approach by considering
the influence of fields with the boundary on the particle. The possible
observational effects will be discussed.

\section{Influence functional approach}
We consider the dynamics of a nonrelativistic point particle of
charge $e$ interacting with quantized electromagnetic fields. In the Coulomb gauge, $\mathbf{\nabla}\cdot \mathbf{A}=0$, the Lagrangian is expressed as
\begin{eqnarray}\label{lagrangian-charge}
    L[\mathbf{q},\mathbf{A}_{\mathrm{T}}]&=\frac{1}{2}m\dot{\mathbf{q}}^2-V(\mathbf{q})-\frac{1}{2}
    \int\!d^3\mathbf{x}\,d^3\mathbf{y}\;\varrho(x;\mathbf{q})G(\mathbf{x},\mathbf{y})\varrho(y;\mathbf{q})\nonumber\\
    &\qquad+\int\!d^3\mathbf{x}\;\left[\frac{1}{2}(\partial_\mu\mathbf{A}_{\mathrm{T}})^2+\mathbf{j}\cdot\mathbf{A}_{\mathrm{T}}\right]\,,
\end{eqnarray}
in terms of the transverse components of the gauge potential $\mathbf{A}_{\mathrm{T}}$, and the position $\mathbf{q}$ of the point charge. The instantaneous Coulomb Green's function satisfies the Gauss's law, meanwhile the charge and current densities take the form
\begin{equation}
\varrho (x; {\bf q}(t)) =e\, \delta^{(3)} ( {\bf x}-{\bf q}(t) ) \,,\qquad
{\bf j} (x;  {\bf q}(t)) = e\,  \dot{\bf q} (t) \, \delta^{(3)} ( {\bf
x}-{\bf q}(t) ) \, . \label{charge-current}
\end{equation}
Let ${\hat \rho}(t)$ be the density matrix of the particle-field
system, and then it evolves unitarily according to
\begin{equation}
	\hat{\rho}(t_f) = U(t_f, t_i) \, {\hat \rho} (t_i) \, U^{-1} (t_f,
t_i )
\end{equation}
with $ U(t_f,t_i) $ the unitary time-evolution operator.
It is convenient to assume that the state of the particle-field at
an initial time $t_i$ is facterizable as ${\hat \rho} (t_i) = {\hat
\rho}_{e} (t_i) \otimes {\hat \rho}_{{\bf A}_{\rm T}} (t_i)$. The
more sophisticated scheme of the density matrix involving initial
correlations can be found in Ref.~\cite{GR}.
The gauge field at the time $t_i$ is assumed in  the vacuum state.
Since the coupling between the electron and the fields is linear, the field variables can be traced out exactly. Thus, this influence functional takes full account of the backreaction from electromagnetic fields. The physics becomes more
transparent when we write the evolution of the reduced density
matrix in the following form
\begin{equation}
    \rho_r(\mathbf{q}_f,\tilde{\mathbf{q}}_f,t_f)=\int\!d^3\mathbf{q}_1\,d^3\mathbf{q}_2\;\mathcal{J}(\mathbf{q}_f,\tilde{\mathbf{q}}_f,t_f;\mathbf{q}_1,\mathbf{q}_2,t_i)\,\rho_{e}(\mathbf{q}_1,\mathbf{q}_2,t_i)\,,\label{evolveelectron}
\end{equation}
where the propagating function
$\mathcal{J}(\mathbf{q}_f,\tilde{\mathbf{q}}_f,t_f;\mathbf{q}_1,\mathbf{q}_2,t_i)$
is
\begin{equation}\label{propagator}
    \mathcal{J}=\int^{\mathbf{q}_f}_{\mathbf{q}_1}\!\!\mathcal{D}\mathbf{q}^+\!\!\int^{\tilde{\mathbf{q}}_f}_{\mathbf{q}_2}\!\!\mathcal{D}\mathbf{q}^-\;\exp\left[i\int_{t_i}^{t_f}dt\;\Bigl(L_{e}[\mathbf{q}^+]-L_{e}[\mathbf{q}^-]\Bigr)\right]
    \mathcal{F}[\,\mathbf{j}^+_{\mathrm{T}},\mathbf{j}^-_{\mathrm{T}}]\,,
\end{equation}
and the electron Lagrangian $L_e[\mathbf{q}]$ is given by
\begin{equation}\label{lag}
    L_e\bigl[\mathbf{q}\bigr]=\frac{1}{2}m\dot{\mathbf{q}}^2-V(\mathbf{q})-\frac{1}{2}\int\!d^3\mathbf{x}\,d^3\mathbf{y}\;\varrho(x;\mathbf{q})\,G(\mathbf{x},\mathbf{y})\,\varrho(y;\mathbf{q})\,.
\end{equation}
Here we introduce the influence functional $\mathcal{F}[\,\mathbf{
j}^+_{\mathrm{T}},\mathbf{j}^-_{\mathrm{T}}]$,
\begin{eqnarray}\label{ss}
    \mathcal{F}[\,\mathbf{j}^+,\mathbf{j}^-]=&\exp\biggl\{-\frac{1}{2}\,\int d^4x\!\!\int\!d^4x'\Bigl[\nonumber\\
    &\quad\,{\mathbf{j}^+}_i(x;\mathbf{q}^+(t))\,\bigl<{\mathbf{A}^+_{\mathrm{T}}}^i(x){\mathbf{A}^+_{\mathrm{T}}}^j(x')\bigr>\,{\mathbf{j}^+}_j(x';\mathbf{q}^-(t'))\Bigr.\biggr.\nonumber\\
    &-{\mathbf{j}^+}_i(x;\mathbf{q}^+(t))\,\bigl<{\mathbf{A}^+_{\mathrm{T}}}^i(x){\mathbf{A}^-_{\mathrm{T}}}^j(x')\bigr>\,{\mathbf{j}^-}_j(x';\mathbf{q}^-(t'))\nonumber\\
    &-{\mathbf{j}^-}_i(x;\mathbf{q}^-(t))\,\bigl<{\mathbf{A}^-_{\mathrm{T}}}^i(x){\mathbf{A}^+_{\mathrm{T}}}^j(x')\bigr>\,{\mathbf{j}^+}_j(x';\mathbf{q}^+(t'))\nonumber\\
    \biggl.\Bigl.&+{\mathbf{j}^-}_i(x;\mathbf{q}^-(t))\,\bigl<{\mathbf{A}^-_{\mathrm{T}}}^i(x){\mathbf{A}^-_{\mathrm{T}}}^j(x')\bigr>\,{\mathbf{j}^-}_j(x';\mathbf{q}^-(t'))\Bigr]\biggr\}\,,\label{influencefun}
\end{eqnarray}
which contains full information about the backreaction effects of quantized
electromagnetic fields on the electron, and is a highly nonlocal
object. The Green's functions of the vector potential are defined by
\begin{eqnarray}
    \bigl<{\mathbf{A}^+_{\mathrm{T}}}^i(x){\mathbf{A}^+_{\mathrm{T}}}^j(x')\bigr>&=&\bigl<{\mathbf{A}_{\mathrm{T}}}^i(x){\mathbf{A}_{\mathrm{T}}}^j(x')\bigr>\,\theta(t-t')+\bigl<{\mathbf{A}_{\mathrm{T}}}^j(x'){\mathbf{A}_{\mathrm{T}}}^i(x)\bigr>\,\theta (t'-t)\,,\nonumber\\
    \bigl<{\mathbf{A}^-_{\mathrm{T}}}^i(x){\mathbf{A}^-_{\mathrm{T}}}^j(x')\bigr>&=&\bigl<{\mathbf{A}_{\mathrm{T}}}^j(x'){\mathbf{A}_{\mathrm{T}}}^i(x)\bigr>\,\theta(t-t')+\bigl<{\mathbf{A}_{\mathrm{T}}}^i(x){\mathbf{A}_{\mathrm{T}}}^j(x')\bigr> \,\theta (t'-t)\,,\nonumber\\
    \bigl<{\mathbf{A}^+_{\mathrm{T}}}^i(x){\mathbf{A}^-_{\mathrm{T}}}^j(x')\bigr>&=&\bigl<{\mathbf{A}_{\mathrm{T}}}^j(x'){\mathbf{A}_{\mathrm{T}}}^i(x)\bigr>\equiv\mathrm{Tr}\left\{\rho_{\mathbf{A}_{\mathrm{T}}}\,{\mathbf{A}_{\mathrm{T}}}^j(x'){\mathbf{A}_{\mathrm{T}}}^i(x)\right\}\,,\nonumber \\
    \bigl<{\mathbf{A}^-_{\mathrm{T}}}^i(x){\mathbf{A}^+_{\mathrm{T}}}^j(x')\bigr>&=&\bigr<{\mathbf{A}_{\mathrm{T}}}^i(x){\mathbf{A}_{\mathrm{T}}}^j(x')\bigr>\equiv\mathrm{Tr}\left\{\rho_{\mathbf{A}_{\mathrm{T}}}\,{\mathbf{A}_{\mathrm{T}}}^i(x){\mathbf{A}_{\mathrm{T}}}^j(x')\right\}\,,\label{noneqgreenfun}
\end{eqnarray}
and can be explicitly constructed. In particular, the retarded
Green's function and Hadamard function of  vector potentials are
defined respectively by
\begin{eqnarray}
    G_{R}^{ij}(x-x')&=&i\,\theta(t-t')\,\bigl<\left[{\mathbf{A}_{\mathrm{T}}}^i(x),{\mathbf{A}_{\mathrm{T}}}^j(x')\right]\bigr>\,,\label{commutator}\\
    G_{H}^{ij}(x-x')&=&\frac{1}{2}\,\bigl<\left\{{\mathbf{A}_{\mathrm{T}}}^i(x),{\mathbf{A}_{\mathrm{T}}}^j(x')\right\}\big>\,.\label{anticommutator}
\end{eqnarray}

In the presence of the perfectly conducting plate, the tangential
component of the electric field $\mathbf{E}$ as well as the normal
component of the magnetic field $\mathbf{B}$ on the plate surface
vanish. When the plate is placed at the $z=0$ plane, the transverse
vector potential $\mathbf{A}_{\mathrm{T}}$ in the $z>0$ region is
given by~\cite{BA1, BA2},
\begin{eqnarray}
    && \mathbf{A}_{\mathrm{T}}(x)=\int\!\frac{d^2\mathbf{k}_{\parallel}}{2\pi}\!\int_0^{\infty}\!\frac{dk_z}{(2\pi)^{1/2}}\;\frac{2}{\sqrt{2\omega}}\biggl\{a_1(\mathbf{k})\,\hat{\mathbf{k}}_{\parallel}\times\hat{\mathbf{z}}\,\sin k_zz\biggr.\nonumber \\
 &&\biggl.+\;a_2(\mathbf{k})\left[i\,\hat{\mathbf{k}}_{\parallel}\left(\frac{k_z}{\omega}\right)\sin k_zz -\hat{\mathbf{z}}\left(\frac{k_{\parallel}}{\omega}\right)\cos k_zz\right]\biggr\}\;e^{i\mathbf{k}_{\parallel}\cdot\mathbf{x}_{\parallel}-i\omega
 t}+\rm{H.C.}\,,
\end{eqnarray}
where the circumflex identifies unit vectors. The position vector
$\mathbf{x}$ is decomposed into $\mathbf{x}=( \mathbf{x}_{\parallel},z)$
where $\mathbf{x}_{\parallel}$ is the components parallel to the
plate. Similarly, the wave vector is expressed by  $\mathbf{k}=(
\mathbf{k}_{\parallel},k_ z)$ with $\omega^2=k_{\parallel}^2+k_z^2$.
The creation and annihilation operators obey the typical commutation
relations of the free fields.

\section{Decoherence induced by vacuum
fluctuations with the boundary }\label{sec2}

First, we study the decoherence dynamics of the electron coupled to
quantized electromagnetic fields in the presence of the conducting
plate. Let us now consider the initial electron state vector
$\bigl|\Psi(t_i)\bigr>$ to be a coherent superposition of two
localized states $\bigl|\psi_1\bigr>$ and $\bigl|\psi_2\bigr>$ along worldlines $\mathcal{C}_1$ and
$\mathcal{C}_2$, respectively, after they leave the beam splitter at
the moment $t_i$,
\begin{equation}
    \bigl|\Psi(t_i)\bigr>=\bigl|\psi_1(t_i)\bigr>+\bigl|\psi_2(t_i)\bigr>\,.
\end{equation}
As such, the leading effect of the decoherence
can be obtained by evaluating the propagating function
(\ref{propagator}) along prescribed classical paths of the
electrons. Thereby, the diagonal components of the reduced density
matrix $\rho_r(\mathbf{q}_f,\mathbf{q}_f,t_f)$ now becomes
\begin{eqnarray}
    \rho_r(\mathbf{q}_f,\mathbf{q}_f,t_f)&=\bigl|\psi_1(\mathbf{q}_f,t_f)\bigr|^2+\bigl|\psi_2(\mathbf{q}_f,t_f)\bigr|^2\nonumber \\
    & \quad\quad+2\,e^{\mathcal{W}[\,\bar{\mathbf{j}}^1,\bar{\mathbf{j}}^2]}\;\mathrm{Re}\left\{e^{i\,\Phi[\,\bar{\mathbf{j}}^1,\bar{\mathbf{j}}^2]}\psi_1^{\vphantom{*}}(\mathbf{q}_f,t_f)\,\psi_{2}^{*}(\mathbf{q}_f,t_f)\right\}\,,
\end{eqnarray}
where the $\mathcal{W}$ and $\Phi$ functionals are found to be
\begin{eqnarray}
    \Phi[\,\bar{\mathbf{j}}^{1},\bar{\mathbf{j}}^{2}]& =\frac{1}{2}\,\!\!\int\!d^4x\!\!\int\!d^4x'
    \Bigl[\,\bar{\mathbf{j}}^{1}_{i}(x;\mathbf{q}^1)-\bar{\mathbf{j}}^{2}_{i}(x;\mathbf{q}^2)\Bigr]
    \nonumber\\
    & \quad\quad\quad\quad\quad\quad\quad\times G_{R}^{ij}(x-x')\Bigl[\,\bar{\mathbf{j}}_{j}^{1}(x';\mathbf{q}^1)+\bar{\mathbf{j}}^{2}_{j}(x';\mathbf{q}^2)\Bigr]\,,\nonumber\\
    \mathcal{W}[\,\bar{\mathbf{j}}^{1},\bar{\mathbf{j}}^{2}]&=-\frac{1}{2}\,\!\!\int\!d^4x\!\!\int\!d^4x' \Bigl[\,\bar{\mathbf{j}}^{1}_{i}(x;\mathbf{q}^1)-\bar{\mathbf{j}}^{2}_{i}(x;\mathbf{q}^2)\Bigr] \nonumber\\
    &  \quad\quad\quad\quad\quad\quad\quad\times G_{H}^{ij}(x-x')\Bigl[\,\bar{\mathbf{j}}^{1}_{j}(x';\mathbf{q}^1)-\bar{\mathbf{j}}^{2}_{j}(x';\mathbf{q}^2)\Bigr]\,.\label{phase-decoherence}
\end{eqnarray}
Here $\bar{\mathbf{j}}^{1,2}$ is the classical current
along the respective paths,  $\mathcal{C}_1$ and $\mathcal{C}_2$ to
be specified later. The evolution of the electron states
$\psi_{1,2}(\mathbf{q}_f,t_f)$ is governed by the Lagrangian $L_e$
in Eq.~(\ref{lag}) with backreaction effects neglected. Under the classical approximation with the prescribed electron's
trajectory dictated by an external potential $V$, we find that the
exponent of the modulus of the influence functional describes the
extent of the amplitude change of interference contrast, and is
determined by the Hadamard function of vector potentials. Its
phase results in an overall shift for the interference pattern, and is
related to the retarded Green's function.

The path plane on which the electrons travel can be either parallel
or perpendicular to the plate. When the path plane is normal to the
conducting plate, the electron worldlines are given by
$\mathcal{C}_{1,2}=(t,v_xt,0,z_0\pm\zeta(t))$. We will choose a
frame $\mathfrak{S}$ which moves along the worldline
$(t,v_xt,0,z_0)$ and has the same orientation as the laboratory
frame. In this frame, the electrons are seen to have sideways motion
in the $z$ direction only. Then the $\mathcal{W}_{\perp}$ functional
depends on the $z$--$z$ component of the vector potential Hadamard
function. Under the dipole approximation, the corrections to the
decoherence functional due to the presence of the conducting plate
is expressed in terms of the ratio $\xi=z_0/T$ with $z_{0}$ being the effective distance of the
electrons to the plate and $2T$ the effective flight time.
Asymptotically, the ratio
$\mathcal{W}_{\perp}/\left|\mathcal{W}_0\right|$ is given by
\begin{equation}\label{ds}
    \frac{\mathcal{W}_{\perp}}{\left|\mathcal{W}_0\right|}=\cases{
                                                                -2+\displaystyle\frac{8}{5}\xi^2+\mathcal{O}(\xi^4)\,, & $\xi\to 0 $\,;\\
                                                                &\\
                                                                -1-\displaystyle\frac{3}{16}\frac{1}{\xi^4}+\mathcal{O}(\frac{1}{\xi^{6}})\,,
                                                                &$ \xi\to\infty $ \,,}
\end{equation}
Here the $\mathcal{W}_0$ functional denotes the contribution solely from unbounded Minkowski vacuum.

On the other hand, when the path plane lies parallel to the
conducting plate, here the electron worldlines are given by
$\mathcal{C}_{1,2}=(t,v_xt,\pm\zeta(t),z_0)$.  The same reference
frame $\mathfrak{S}$ is chosen so that the electrons are seen to
move in the $y$ direction. Then, the $y$--$y$ component of the
vector potential Hadamard function becomes relevant to the
$\mathcal{W}_{\parallel}$. Following the same approximation, asymptotically
$\mathcal{W}_{\parallel}/\left|\mathcal{W}_0\right|$ is obtained as
\begin{equation}
    \frac{\mathcal{W}_{\parallel}}{\left|\mathcal{W}_0\right|}=\cases{
                                                                -\displaystyle\frac{16}{5}\xi^2+\frac{144}{35}\xi^4+\mathcal{O}(\xi^6)\,, &$ \xi\to 0 $\,;\\
                                                                &\\
                                                                -1-\displaystyle\frac{3}{16}\frac{1}{\xi^4}+\mathcal{O}(\frac{1}{\xi^6})\,, &$ \xi\to\infty $\,.}
\end{equation}
It is found that the effects of coherence reduction of the electrons
by zero-point fluctuations with the boundary are strikingly deviated
from that without the boundary. Thus, the presence of the conducting
plate anisotropically modifies electromagnetic vacuum fluctuations
that in turn influence the decoherence dynamics of the electrons. Electron
coherence is enhanced when the path plane of the
electrons is parallel to the plate. This results from the
suppression of zero-point fluctuations due to the boundary condition
in the direction parallel to the plate. On the other hand, the
electron coherence is reduced in the perpendicular configuration
where zero-point fluctuations are boosted along the direction normal
to the plate.

\section{Stochastic dynamics of a point charge driven by electromagnetic vacuum
fluctuations in the presence of the conducting plate}

The anisotropy of electromagnetic vacuum fluctuations in the
presence of the conducting plate has been studied via an
interference experiment of the electrons, and is manifested in the
form of the amplitude change and phase shift of the interference
fringes~\cite{FO,JT1}. Here we wish to further explore
the anisotropic nature of electromagnetic vacuum fluctuations by the
motion of the charged particle~\cite{YU,JT3}. We  assume that the
particle initially is in  a localized state, and thus its density matrix can be
expanded by the position eigenstate of the eigenvalue
$\mathbf{q}_i$,
\begin{equation}
{\hat \rho}_{e} (t_i) = \left|\mathbf{q}_i,t_i \right>\left<
\mathbf{q}_i, t_i \right|\,. \label{initialcondition_e}
\end{equation}
In Eq.~(\ref{ss}), it is found more convenient to change the variables $\mathbf{q}^{+}$
and $\mathbf{q}^{-}$ to the average and relative coordinates, $\mathbf{q}=\left(\mathbf{q}^+ +\mathbf{q}^-\right)/2$ and $\mathbf{r}=\mathbf{q}^+ -\mathbf{q}^-$. Next, we
introduce the auxiliary noise fields $\xi^i (t)$ with the Gaussian
distribution function,
\begin{equation}
    \mathcal{P}[\xi^i(t)]=\exp\left\{-i\frac{\hbar}{2}\int_{-\infty}^{\infty}\!dt\!\int_{-\infty}^{\infty}\!dt'\;\xi^i(t)\,G_{H}^{ij}{}^{-1}\left[{\bf q}(t),{\bf q}(t');t-t'\right]\,\xi^j(t')\right\}
\label{noisedistri}
\end{equation}
and thus the imaginary part of the coarse-grained action can be
expressed as a functional integration over $\xi_i (t)$ weighted by
the distribution function $ \mathcal{P}[\xi_i (t)]$. As a result, we
end up with
\begin{eqnarray}
    \exp\left\{\frac{i}{\hbar}S_{CG}[\mathbf{q},\mathbf{r}]\right\}=\int\mathcal{D}
\xi_i \; &\mathcal{P} [\xi_i (t)] \, \exp\left\{\frac{i}{\hbar} \biggl[ \mathrm{Re}\{S_{CG}\left[ {\bf q} , {\bf r} \right] \}\biggr.\right.\nonumber\\
		&\left.\biggl.-\;\hbar\,e \int_{-\infty}^{\infty}dt\; r^i \left( \delta^{ij}\frac{d}{d t} - q^j(t)\nabla^i \right)\xi^j \biggr]\right\} \,. \label{effectaction}
\end{eqnarray}
The expressions in the squared brackets on the right hand side is
defined as the stochastic effective action, which consists of the
real part of the coarse-grained effective action as well as a
coupling term of the relative coordinate $r^i$ with the stochastic
noise $\xi^i$. The  Langevin equation is obtained by extremizing the
stochastic effective action and then setting $r^{i}$ to zero. By
doing so, we have ignored intrinsic quantum fluctuations of the
particle, and that holds as long as the resolution of the
measurement on length scales is greater than its position
uncertainty. The Langevin equation is then given by~\cite{JT3}
\begin{eqnarray}
	m \ddot{q}^i &+ \nabla^i V(\mathbf{q}(t)) + e^2\nabla^iG [{\bf q}(t),{\bf q}(t)]\nonumber \\
&+e^2\left(\delta^{i l} \frac{d}{dt} - \dot{q}^l (t) \nabla_i\right)\int_{-\infty}^{\infty}dt' \; G_{R}^{lj} \left[{\bf q}(t),{\bf q}(t'); t-t'\right]\,\dot{q}^j (t') \nonumber\\
&\qquad\qquad\qquad= -\hbar\, e \, \left( \delta^{il} \frac{d}{d t} - \dot{q}^l (t)\nabla^i \right) \, \xi^l (t)\label{nonlinearlangevin}
\end{eqnarray}
with the noise-noise correlation functions,
\begin{equation}
\langle\xi^i(t)\rangle =0\,,\qquad\qquad\langle\xi^i (t)\xi^j(t')\rangle=\frac{1}{\hbar} G_{H}^{ij} \left[{\bf q}(t), {\bf q}(t');
t-t'\right]\, . \label{noisecorrel}
\end{equation}
This Langevin equation encompasses fluctuations and
dissipation effects on the charge's motion from quantized
electromagnetic fields via the kernels $G_{H}^{ij}$ and $G_{R}^{ij}$
respectively, which both are in turn linked by the
fluctuation-dissipation relation~\cite{KU}.

The backreaction kernel function of  electromagnetic fields appears
purely classical due to the fact that the coupling between a point
charge and electromagnetic gauge potentials is linear. The
noise-noise correlation functions can in principle be computed by
taking an appropriate statistical average with the distribution
functional $\mathcal{P}[\xi^{i}(t)]$.
The noise-averaged result that describes the dynamics of the mean trajectory
reduces, in the case of free space, to the known
Abraham-Lorentz-Dirac equation with the self-force~\cite{HU1}.
In the presence of the boundary, it can be seen that radiation emitted by the charge in nonuniform motion should be bounced back and then impinge upon the charge at later
times. This will give rise to an additional retardation effect,
which in turn results in a non-Markovian evolution of the charged particle.
The fluctuations off the mean
trajectory are driven by the stochastic noise, and will be studied with
backreaction dissipation taken into consideration in a way that a
underlying fluctuation-dissipation relation is obeyed.
The noise-driven trajectory fluctuations thus are entirely of quantum
origin as can be seen from an explicit $\hbar$ dependence
in the noise term as well as the noise-noise correlation.
As it stands, this is a nonlinear Langevin equation with
non-Markovian backreaction, and the noise depends in a complicated
way on the charge's trajectory because the noise correlation
function itself is a functional of the trajectory.

The integro-differential equation~(\ref{nonlinearlangevin}) can be
cast into a form similar to the Lorentz equation~\cite{JT3}. We
consider that a charged particle undergoes a harmonic
motion, and assume that the amplitude of oscillation is sufficiently
small. The appropriate approximation for a non-relativistic motion
will be the dipole approximation. This approximation amounts to
considering the backreaction solely from electric fields, and
linearizing the Langevin equation in such a way that the equation of
motion remains non-Markovian. The ultraviolet
divergence arises due to the integration of backreaction effects
from the free-space contribution over all energy scales of fields in
the coincidence limit.  The energy cutoff $\Lambda$ is
then introduced to regularize the integral. The cutoff scale can
be chosen to be the inverse of the width of the electron wavefunction. It
essentially quantifies the intrinsic uncertainty on the charged
particle. The divergence is then absorbed into mass renormalization.

The velocity fluctuations of the charged oscillator near the boundary are found to grow linearly with time in the early stage of the evolution, and then are asymptotically saturated as a
result of the fluctuation-dissipation relation~\cite{JT3}. From dimensional consideration, the saturated value of
velocity fluctuations induced by the presence of the boundary is given
by $\Delta v^2_{B}\sim e^2/ m_e^2 z_0^2 $, while that arising from
the electron's motion can be argued to be $\Delta
v^2_{M}\sim\overline{\omega}_0/m_e $. Then, the ratio of two effects
is
\begin{equation}
    \frac{\Delta v^2_{M}}{\Delta v^2_{B}}\sim 10^7 \, \left(\frac{\overline{\omega}_0}{10^{12}\,\mathrm{s}^{-1}}\right) \,\left(\frac{z_0}{\mu\mathrm{m}}\right)^2 \, .
\end{equation}
 As a result,
 velocity fluctuations owing to the electron's motion are
overwhelmingly dominant if its distance to the plate $z_0$  is about the
order of $\mu\mathrm{m}$, and the oscillation frequency, say
$10^{12}\,\mathrm{s}^{-1}$, is chosen below the plasma frequency of the plate
$10^{16}\,\mathrm{s}^{-1}$~\cite{JD}.
The
corresponding modification in the effective temperature is
\begin{equation}
    T_{eff}\sim\frac{\hbar\overline{\omega}_0}{k_{B}}\sim10\left(\frac{\overline{\omega}_0}{10^{12}\,\mathrm{s}^{-1}}\right)\,{\rm K}\,,
\end{equation}
where $k_B$ is the Boltzmann constant.

\section{Concluding remarks}
The nature of electromagnetic vacuum fluctuations in the presence of the conducting plate
is studied by its effects on charged particles.
The aim of this paper is to present a field-theoretic approach by considering the influence of fields with the boundary on the particle with the
method of Feynman-Vernon influence functional.
The extension of the above study by involving thermal fluctuations within the same formalism is in progress.

\section*{Acknowledgments}
This work was supported in part by the
National Science Council, R. O. C. under grant
NSC95-2112-M-259-011-MY2.

\end{document}